\documentclass[prb,twocolumn,aps]{revtex4-1}
\usepackage{graphicx}
\usepackage{dcolumn}
\usepackage{bm}
\usepackage{hyperref}

\newcommand{\beq}{\begin{eqnarray}}
\newcommand{\eeq}{\end{eqnarray}}

\begin{document}\title{ Emergence of Particle-Hole Symmetry near Optimal Doping
in High-Temperature Copper Oxide Superconductors}
\author{\small Shiladitya Chakraborty }
\author{Dimitrios Galanakis}
\author{Philip Phillips}
\affiliation{Department of Physics,
University of Illinois
1110 W. Green Street, Urbana, IL 61801, U.S.A.}
\date{\today}

\begin{abstract}
   High-temperature copper oxide superconductors (cuprates) display unconventional
 physics when they are lightly doped whereas the
 standard theory of metals prevails in the opposite regime.  For example, the thermoelectric power, that is the voltage that develops across
 a sample in response to a temperature gradient, changes
 sign abruptly near optimal doping in a wide class of cuprates, a stark departure from the
 standard theory of metals in which the thermopower vanishes only when
 one electron exists per site.   We show that this effect arises from
  proximity to a state in which particle-hole symmetry is dynamically
 generated. The operative mechanism is dynamical spectral weight
  transfer from states that lie at least 2eV away from the chemical
  potential.   We show that the sign change is reproduced quantitatively
  within the Hubbard model for moderate values of the on-site
  repulsion, $U$.  For sufficiently large values of on-site repulsion,
  for example, $U=20t$, ($t$ the hopping matrix element), dynamical
  spectral weight transfer attenuates and our calculated results for
  the thermopower are
  in prefect agreement with exact atomic limit. The emergent particle-hole symmetry close to optimal doping
  points to pairing in the cuprates being driven by high-energy
  electronic states. 
\end {abstract}
 
\maketitle

\section{Introduction}
 
As charge carriers are doped into the wide family of parent copper-oxide ceramics
  (cuprates), superconductivity obtains with a maximum transition
  temperature at a particular chemical composition.  
  The result is a dome-shaped superconducting region in the doping-temperature
  plane.   The
  origin of the dome remains pivotal to the solution of this problem
  because the properties of all high $T_c$ superconductors change
  drastically around optimal doping, $x_c$.  Unconventional physics
  pervades for $x<x_c$ in which the standard theory of metals
  breaks down whereas such a description is recovered in the opposite
  regime.  The efficient cause underlying this drastic change in the
  physics of the cuprates as the dome is traversed remains one of the
  key mysteries of these materials.  To solve this problem, it is instructive to
focus on a correlate of superconductivity, that is a phenomenon which
supervenes on superconductivity and exhibits an abrupt signature at the
top of the dome. Such a correlate of superconductivity could elucidate
the ubiquitous dome-like structure in the doping-temperature plane of
the cuprates.    

To this end, we consider the thermo-electric power ($S$), defined as the voltage
difference that develops across a material in response to a
temperature gradient. 
 Physically, the thermopower measures the
entropy per carrier and is a direct probe of particle number
rather than charge conservation, a distinction which is of utmost
importance in this work.  A universal feature observed amongst different
families of hole-doped cuprates\cite{honma,matsuura,obertelli,cooper2,greene} is that
the thermo-electric power is non-zero everywhere except at 
one particular doping value.  Shown in Fig. (\ref{tpexp}a) is a
collation of thermopower data at 290K
for a wide range of hole-doped cuprates as a function of the planar
hole density per unit area, $P_{\rm pl}$, which measures the hole
concentration in each CuO$_2$ layer.  $P_{\rm pl}$ provides an
even-handed way of comparing different families of cuprates that is
independent of the doping method or sample quality, be they single
crystals or not.  As long as $x$ refers to the number of holes per
CuO$_2$ layer, $x$ and $P_{\rm pl}$ can be used interchangeably as we
do here.  Clearly shown is that for all
 hole-doped cuprates, the thermopower at 290K is positive for $P_{\rm pl}<0.24$,
whereas for $P_{\rm pl}>0.24$, it is negative.  To correlate this sign change
with superconductivity, Honma and Hor\cite{honma} used the thermopower
doping scale as opposed to the Presland scale\cite{thermoboys},
$1-T_c/T_c^{\rm max}=82.6(x-0.16)^2$,
in which the maximum $T_c$ is artificially fixed to be $0.16$, to
determine the doping level at which $T_c$ is maximized.  We show in
Fig. (\ref{tpexp}b)
the maximum superconducting transition temperature $T^{\rm max}_c$ as a function
of $P_{\rm pl}$ for 23 different types of cuprates classified by the
number of copper-oxide layers in each unit cell.  Of the 23 cuprates
shown, only four single-layer materials exhibit any significant deviation of the maximum $T_c$ from the planar
hole density at which the thermopower vanishes.

\begin{widetext}
\begin{figure}
\includegraphics[width=15.5cm]{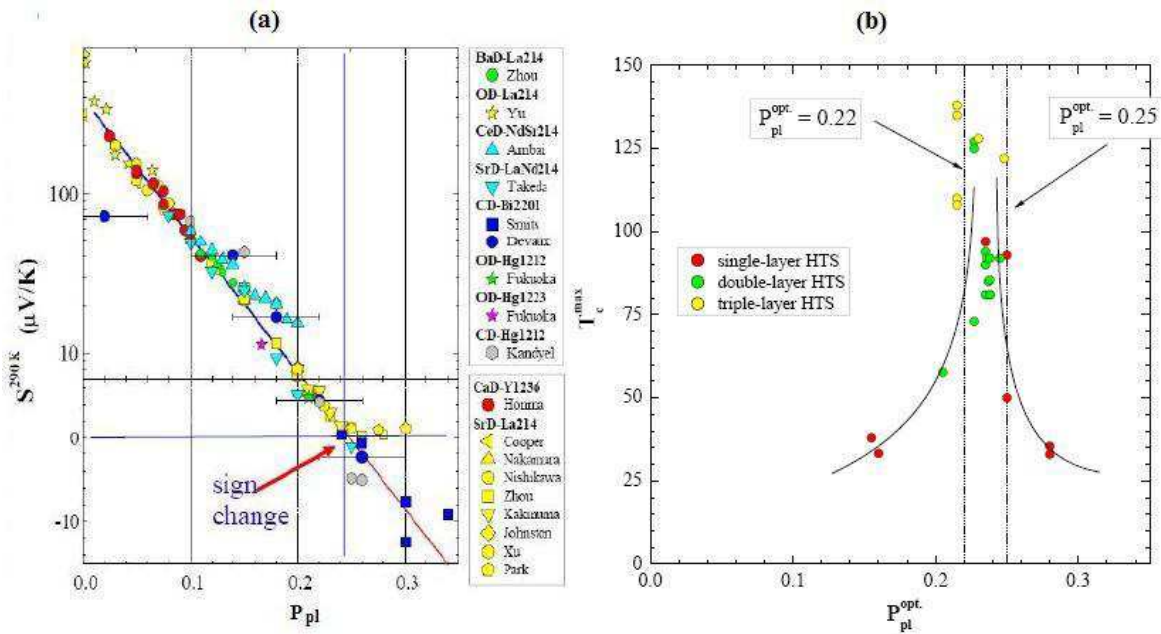}
\caption{ a) Universal behavior of  the thermoelectric power\cite{honma} (290K) as a
 function of planar hole density ($P_{\rm pl}$), for various families
of hole-doped
 cuprates.   All exhibit a sign change at $P_{\rm pl}=0.23$. Above the solid-bold horizontal line, the thermpower obeys the
 functional form, $S^{290}(P_{\rm pl})=392\exp(-19.7P_{\rm pl})$ for
 $0.01<P_{\rm pl}<0.21$.  Below the solid-bold horizontal line,
 $S^{290}(P_{pl})=40.47-163.4P_{\rm pl}$ for $0.21<P_{\rm pl}<0.34$.
 These functional forms were used\cite{honma} to determine the hole doping levels
 for all the cuprates rather than the widely used empirical formula\cite{thermoboys}
 $1-T_c/T_c^{\rm max}=82.6(x-0.16)^2$  which artificially fixes the
 optimal doping level of all cuprates to be $0.16$.  b)
 Maximum transition temperature as a function of the planar hole
 density using the thermopower scale to determine the doping level.  Except for three single-layer materials, the vanishing of
 the thermopower coincides with the doping level
 at which the transition temperature is maximized.}
\label{tpexp}
\end{figure}

One might argue that
this agreement is irrelevant since the thermopower is highly
temperature dependent.  However, experiments\cite{honma2} indicate that at 110K,
the zero-crossing of the thermopower is shifted to $0.26$, quite
similar to the value at $290K$.  In addition, one might also question
the validity of the thermopower scale to calibrate the doping level.
To address this question,  we reprint here a figure
(see Fig. (\ref{honmafig}))
from their paper which illustrates that the thermopower scale
 is in excellent agreement with the doping level in Y123
determined by three different experimental methods.  Hence, the zero-crossing in the
vicinity of $T_c^{\rm max}$ is a robust feature of the cuprates that
has direct bearing on why $T_c$ is so high.  Our analysis indicates that mixing with states that lie
at least $2eV$  (the optical gap in the
cuprates\cite{tanner,cooper,uchida,opt0,opt2,opt3}) away from the chemical potential acount for the
dramatic sign change of the thermopower and hence are the cause of the high transition temperature
for most of the cuprates, particularly those exhibiting $T_c>70K$ (see
Fig. (\ref{tpexp}a).

\begin{figure}
\centering
\includegraphics[height=10.0cm]{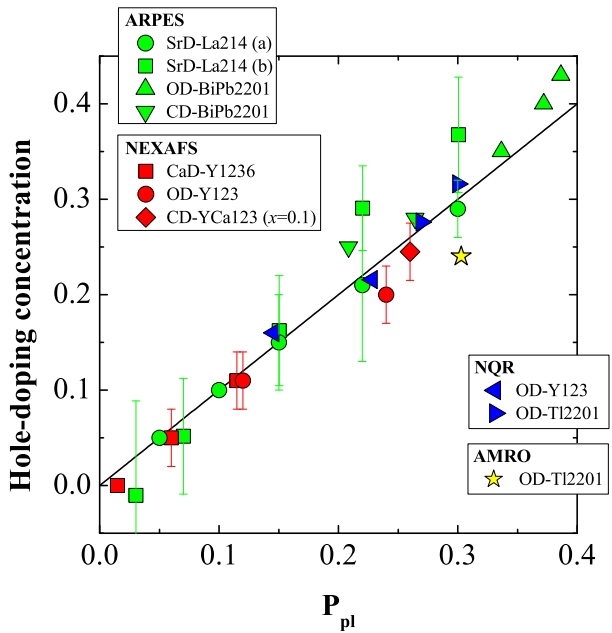}
\caption{Hole-doping level from various techniques compared with the
  doping scale extracted from the thermopower, $P_{\rm pl}$. The red
  points are obtained from (near-edge X-ray absorption fine structure)
  NEXAFS: red
  circles are OD-Y123\cite{ody123}, red diamonds are co-doped
  Ca-YaC123 (at x=0.1)\cite{ody123}, red squares are Calcium-doped
  Y1236\cite{ody123}.  The blue points are from nuclear-quadrupole
  resonance (NQR) measurements: left blue arrow is
  OD-Y123\cite{NQR} and the right blue arrow is
  OD-Ti2201\cite{NQR}.  The green points are from ARPES: green
  circles are Strontium-doped La214\cite{srd214a}, green squares are
  also Strontium-doped La214\cite{srd214b}, uprrow, overdoped
  BiPb2201\cite{arp1}, down arrow co-doped
  BiP2201\cite{arp1}.  The star corresponds to angular
  magnetoresistance oscillations (AMRO)\cite{amro}. Reprinted
from Phys. Rev. B {\bf 77}, 84520 (2008).}
\label{honmafig}
\end{figure} 
\end{widetext}

\section{Thermopower and Spectral Weight Transfer}

The dramatic sign change of the thermopower in the cuprates poses a distinct
theoretical problem because it represents a stark departure from
the predictions of free-electron physics.  For example, in the
standard theory of metals, Fermi liquid theory, the sign of the
thermopower unambiguously reveals the sign of the dominant charge
carriers in the material.  Consequently, in the standard theory of
metals, the thermopower only vanishes in a half-filled band where the
number of filled and empty states is equal, the particle-hole
symmetric condition.  
  In the cuprates, the parent materials possess
a half-filled band.  Doping away from half-filling by the introduction
of holes should only
increase the particle-hole asymmetry.   Consequently, it is unexpected
that the thermopower should vanish at $x=0.24$ corresponding to a
filling of $n=0.76$.   Existing treatments of the thermopower in the
cuprates address either the low-doping regime\cite{plee} or invoke special band
structure effects such as Van Hove singularities\cite{mcintosh} and
the shape of the Fermi surface\cite{hilderbrand} to describe the sign
change.  However, given
that the sign change obtains near  optimal doping, the approximate
crossover from strong to weakly interacting physics, the
explanation of the sign change of $S$, must incorporate the physics of
strong correlations. 

To address this question, we focus on the minimal model
that captures the strong correlations of the copper-oxide planes of
the cuprates.  While what constitutes the minimal model for the cuprates can
certainly be debated, it is clear\cite{p2,p22,p3,p4} that regardless of the model, the
largest energy scale arises from doubly occupying the copper
$d_{x^2-y^2}$ orbital.  This orbital can hybridise with the in-plane
$p_x$ and $p_y$ orbitals and hence a three-band model is natural.   As emphasized earlier\cite{p22}, the hybridization with the d-orbitals in the cuprates is sizeable.  In so far as spectral weight transfer is concerned\cite{p22}, the 3-band model with the p-d hybridization that is
relevant for the cuprates and the one-band Hubbard models are essentially identical for hole doping.  As our primary emphasis here is spectral weight transfer, we adopt the one-band Hubbard model,
\beq\label{hubb}
H_{\rm Hubb}&=&-t\sum_{i,j,\sigma} g_{ij} c^\dagger_{i,\sigma}c_{j,\sigma}+U\sum_{i,\sigma} c^\dagger_{i,\uparrow}c^\dagger_{i,\downarrow}c_{i,\downarrow}c_{i,\uparrow},
\eeq
where $i,j$ label lattice sites, $g_{ij}$ is equal to one iff $i,j$
are nearest neighbours, $c_{i\sigma}$ annihilates an electron with
spin $\sigma$ on lattice site $i$, $t$ is the nearest-neighbour
hopping matrix element and $U$ the energy cost when two electrons
doubly occupy the same site.   Nonetheless, our conclusions carry over naturally to
any n-band model of the cuprates as long as the largest energy scale is
the on-site energy, $U$ in Eq. (\ref{hubb}).  For the cuprates
$U\approx 4.0eV$ and $t\approx 0.5eV$.  Consequently, the cuprates
reside in the strong-coupling regime where standard weak perturbative
treatments breakdown.  

\begin{figure}
\centering
\includegraphics[width=9.0cm]{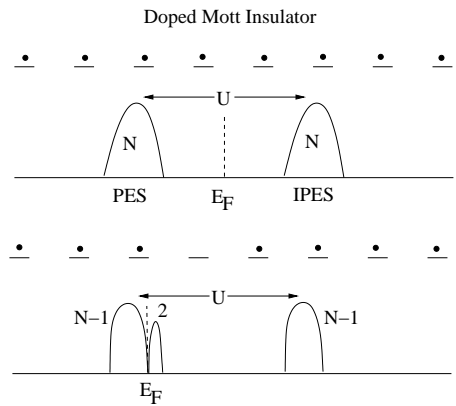}
\caption{Evolution of the single-particle density of states from
  half-filling to the one-hole limit in a doped Mott insulator
  described by the Hubbard model.  Removal of an electron results
  in two empty states at low energy as opposed to one in the
  band-insulator limit. The key difference with the Fermi liquid is
  that the total weight spectral weight carried by the lower Hubbard
  band (analogue of the valence band in a Fermi liquid) is not a
  constant but a function of the filling. }
\label{hubb1}
\end{figure}

Quite generally, the strong correlations in the Hubbard model mediate
a universal sign change in the thermopower at non-integer fillings as
seen in the cuprates.  Since
$U\gg t$ in the cuprates, perturbation around the atomic limit
(vanishing hopping, $t=0$) rather than the non-interacting regime
($U=0$) is the correct starting point.  In 1974,
Beni\cite{beni} considered this limit of the Hubbard model and
calculated the {\bf exact} expression for the thermopower
\beq
S = -\frac{k_{B}}{e}\ln\frac{2x}{1-x}
\label{beni}
\eeq
to $O(t/U)$ and $U\gg k_B T$, where $k_B$ is Boltzmann's constant and
$e$ is the electric charge.  This expression vanishes exactly at a
hole doping level of $x=1/3$.  However, the reason why it vanishes at
$x=1/3$ has never been understood. 
Consider the general expression\cite{mahan2,shastry} for the thermopower ,
\beq\label{Stp}
S = -\frac{k_B}{e}\beta \frac{L_{12}}{L_{11}}, 
\eeq 
with the transport integrals $L_{ij}$ in the relaxation-time approximation given by
\beq
L_{ij} = \displaystyle\int_{-\infty}^{\infty} d\omega(- \frac{\partial f(\omega)}{\partial \omega})\tau ^i(\omega)\omega^{j-1} 
\eeq 
and $\tau(\omega)$ being the relaxation time,
\beq
\tau(\omega) = \frac{1}{N} \displaystyle \sum_{\bf{k},\sigma} (\frac{\partial\epsilon_{\bf k}}{\partial k_x})^2 A^2({\bf k},\omega).
\eeq
In the atomic limit, the single-particle spectral function,
$A(k,\omega)$, which when summed over momentum yields the density of states, has no momentum dependence.  We have used natural units in which $\hbar=m_e=1$. Consequently, from the form of $L_{12}$, a vanishing of the thermopower arises entirely particle-hole symmetry.  How does such a particle-hole symmetry obtain
for $x=1/3$ in a Hubbard system?  Alternatively, what is so special
about the $(2x/1-x)$ ratio in the argument of the logarithm in
Eq. (\ref{beni})?  The answer lies in spectral weight transfer which
was only understood as the signature physical
phenomenon\cite{chen,linio,diag}  of strongly
correlated systems in the early 1990's, roughly 20 years after
Eq. (\ref{beni}) was derived. As no one yet has presented an explanation
of the Beni\cite{beni} result in this light, we offer one here.   In the atomic limit of the Hubbard model, as depicted in
Fig. (\ref{hubb1}), one particle resides per site.  In a Hubbard system
consisting of $N$ sites, there are $N$ ways to remove an electron and
$N$ ways to add one, constituting the photoemission and inverse
photoemission bands, respectively.   In the atomic limit, the splitting
between these bands (also known as the lower and upper Hubbard bands)
is
the Mott gap, $U$, as shown clearly in Fig. (\ref{hubb1}).
Now consider removing a particle.  There are now $N-1$ ways to remove
a particle and $N+1$ ways to add a particle.  However, only $N-1$ such
states lie in the high energy scale as there are only $N-1$ ways to
add a particle to the system so that it costs an energy $U$.  The two
remaining states correspond to the two ways of adding an electron with
either spin up or spin down to the empty site.  Such addition
processes cost no energy and hence must lie directly above the
chemical potential.  In general
if $x$ holes are introduced, the spectral weight immediately above the
chemical potential grows as $2x$ whereas the weight in the lower band
decreases as $1-x$.  The ratio in the Beni\cite{beni} expression is
simply the number of states above and below the chemical potential in
the photoemission band in the atomic limit.  When the two spectral weights are equal, particle-hole
symmetry obtains and the thermopower vanishes.  The exact particle-hole
symmetry condition for the photoemission band of a Mott insulator is
$2x=1-x$.  As a result, in a strongly-correlated system, spectral
weight transfer and the thermopower are intimately linked. 

Real Mott systems are not in the atomic limit because $t/U\ne
0$. This results in two new effects, the second of which destroys the
necessity for a strict particle-hole symmetry to be present for the
thermopower to vanish.  First, the number of
particle addition states immediately above the chemical potential,
\beq
L=\int_\mu^\Lambda N(\omega)d\omega,
\eeq
 computed by integrating the single-particle density of states from
the chemical potential to a cut-off energy that demarcates the low
from the high energy scale, exceeds $2x$\cite{harris}.  The reason is
simple. Away from the atomic limit, no eigenstate of the Hubbard model involves a
fixed number of doubly occupied sites. Harris and Lange\cite{harris}
that such mixing with the high-energy scale leads to an
  increase in the weight of the lower band to $1+x+\alpha$, where
$\alpha$ represents all the correction due to the mixing with the
upper band.  $\alpha$ is strictly positive as shown by Harris and
Lange\cite{harris} and confirmed by numerics\cite{diag}. Because the total weight of both bands is $2$, the weight
of the upper band is $1-x-\alpha$. In general $\alpha$ is doping
dependent and can be as large as 20$\%$\cite{shila2}. To interpret
the thermopower in terms of the entropy, it is helpful to reinstate an atomic-limit picture of the LHB.  We have shown\cite{shila2} how this can be done by re-defining doping level to include the holes that are dynamically generated as a result of mixing with the upper Hubbard band.     To this end, we define $x'=x+\alpha$.  Consequently, the filled part of the spectrum at low energies has a weight of 
$1-x'$ and the empty part $2x'$.   Hence, $L=2x'>2x$ as a result of
dynamical spectral weight transfer. The new particle-hole symmetric condition
 for the lower
band is $1-x-\alpha=2(x+\alpha)$ or $x_{\rm phs}=1/3-\alpha$ which
is strictly less than $1/3$, the atomic-limit value.   However, a strict
particle-hole symmetry is not necessary for the thermopower to vanish
once the hopping is turned on.  At play here is the fact that the spectral function is
momentum dependent for $t\ne 0$. As a result, the vanishing of $L_{12}$ no longer
arises from a simple balancing between the density of states
above and below the chemical potential, but rather $\tau(\omega)$
averaged above and below the chemical potential must be equal.  This pushes
the zero crossing of the thermopower to even lower doping levels.
We note importantly that computational techniques will differ in the
spectral function which in turn will affect the magnitude of the
thermopower.  However, the existence of the zero crossing and its
location relative to $x<1/3-\alpha$ is a robust feature determined
entirely by the integrated weights of the spectral intensity below and
above the chemical potential and not on the computational method used. 

\section{Realistic Calculations}

To quantify the arguments presented here, we
computed Eq. (\ref{Stp}) for the Hubbard model using the cellular-dynamical
mean-field theory\cite{kotliar2}.  In the (CDMFT)method, \cite{kotliar2}
a cluster extending in a small number of sites is treated as the impurity
and therefore the local (cluster) degrees of freedom are treated exactly.
The rest of the lattice, the bath, is described by a multi-component
hybridization function. 

All cluster-DMFT-based algorithms contain the following major components. 

\begin{itemize}
\item An impurity solver, which evaluates the cluster Green function from
the hybridization function.
\item A self consistency condition which expresses the hybridization function
with respect to the cluster Green function.
\item A periodization procedure which connects the lattice quantities with
the cluster quantities.
\end{itemize}
In the present application, we used a 4-site (plaquette) cluster. The
coupling of the cluster to the bath is thus treated in a mean-field
fashion and the cluster quantities are evaluated self consistently
using the cumulant lattice reconstruction 
scheme\cite{stanescukot}. Various impurity solvers
have been proposed in the literature such as exact diagonalization
and quantum Monte-Carlo. However those can only be implemented in
imaginary time and an analytic continuation is required to obtain
real time properties. A real time impurity solver is the non-crossing
approximation (NCA), which is a first order perturbation theory with
respect to the hybridization function. It has the advantage of being
very fast and relatively easy to implement. The NCA has been a valuable
tool for extracting the physics of the Anderson impurity models. The
NCA Equations can be obtained by using the slave boson method \cite{NCAColeman}
and they can be expressed with respect to the pseudo-particle resolvents
and their self energies.  We adopted this technique here in our implementation of the CDMFT method.

\begin{figure}
\includegraphics[width=9cm]{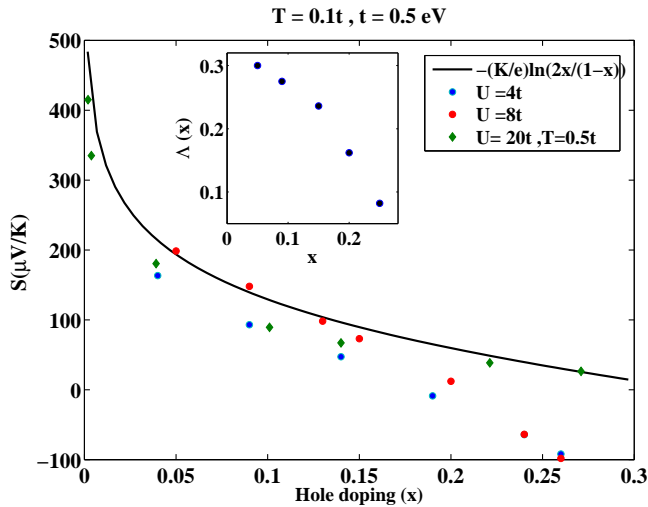}
\caption{Thermoelectric power computed for two different values of
  $U$=8$t$, plotted as a function of hole doping, $x$.  The solid line
  is Eq. (\ref{beni}), the exact result for the thermopower of a doped
Mott insulator in the atomic limit. In the inset $\Lambda$ is the
difference of the spectral weight above and below the chemical
potential.  As is evident, $S$ vanishes at a doping level slightly
lower than that at which $\Lambda=0$.  In a strongly correlated
system, naive particle-hole symmetry is not essential for the
vanishing of the thermopower. }
\label{fig2}
\end{figure}

Using the CDMFT method, we obtained the one-particle spectral function
$A(\bf{k} , \omega)$ for various values of hole doping $\bf{x}$ and
on-site Coulomb repulsion $\bf{U}$ (expressed in units of the nearest
neighbor hopping integral $\bf{t}$) .  To calibrate the CDMFT method
in this context, we
first calculated the thermopower with a large value of $U$ for
comparison purposes with the exact result in the atomic limit of
Beni\cite{beni}. If this method is to be trusted, it should reproduce
these exat results. The diamonds (green in color scale) in
Fig. (\ref{fig2}) represent the thermopower for $U=20t$.  The computed
values match perfectly with the exact asymptotic atomic
limit of Beni\cite{beni}, especially at large dopings.  This striking
agreement indicates
two things.  First, the
method we use here provides an accurate quantitative description of the local physics behind the
spectral weight shifts that leads to the thermopower in the atomic
limit.   Second, by $U=20t$ the dynamical correction to the
spectral weight is negligible.   Consequently, for $U<20t$, the sign
change in the thermopower must occur at $x<1/3$. Indeed this is true.  
From Fig. (\ref{fig2}), we see that both $U=8t$ and $U=4t$ at $T=0.1t$ (slightly higher than the 290K of the experiments) exhibit a sign change before the
atomic limit value of $x=1/3$.  For $U=4t$, the sign change occurs at
$x\approx 0.19$ whereas for $U=8t$, $S$ vanishes at $x=0.21$ in
agreement with the perturbative argument that as $t/U$ decreases, the
critical value of the doping at which the thermopower changes sign
must increase. For $U=8t$, both the magnitude of the thermopower and
the $S=0$ condition are in agreement with the experimental values in
Fig. (\ref{tpexp}a).   The inset shows the difference between the spectral
weight above and below the chemical potential for $U=8t$.  As is
evident, this quantity is finite even when the thermopower vanishes,
in direct constrast to a non-interacting system in which particle-hole
symmetry is essential for $S=0$.  The deviation from
the particle-hole symmetric point, which lies in close proximity to
the doping level at which $S$ vanishes, is due entirely to the
momentum dependence of the spectral function at finite $t/U$.  
Consequently, the vanishing of the thermopower in the 
cuprates at a doping level significantly less than the atomic limit of
$x=1/3$ is a signature that the 
dynamical contributions (the $t/U$ part) to the low-energy spectral
weight and the momentum dependence of the spectral function cannot be
ignored.   Such an 
extreme sensitivity of the thermopower to the spectral weight
redistribution obtains because the thermopower is generated by a
thermal gradient, unlike an electrical gradient in a Hall
measurement.  That is, the thermopower is fundamentally an
experiment about the conservation of particle number (as opposed to
charge conservation in the Hall experiment), which is of
course governed by electron spectral weight shifts.  The final feature
of this story
 
Based on the calculations presented here and our analysis of
Eq. (\ref{beni}), we correlate the vanishing of
the thermopower at the top of the dome for a wide range of cuprates to a proximity to the
 particle-hole symmetric state  which is driven predominantly by
mixing (dynamical spectral weight transfer) with states in the upper
Hubbard band.  As it is the dynamical part of $L$ that is relevant
here, maximising $T_c$  is tied to states at high energy, in
particular states that lie 2eV away from the chemical potential.  This
conclusion is consistent with the results from optical experiments\cite{vdm,btm}
which indicate that the condensation energy in the superconducting
state arises from electronic states that lie $2eV$ away from the chemical
potential. 
 
Aside from influencing the vanishing of the thermopower, dynamical spectral
weight transfer also provides a general mechanism\cite{ftm0}
 for the breakdown of
Fermi liquid theory in doped Mott insulators in the absence of any
symmetry breaking.  In a Fermi liquid, the
phase space available to add a particle at low energies (that is $L$) is exhausted by enumerating
the number of ways ($n_h$) of adding an electron to the empty states.  This
principle follows from the Fermi liquid tenet that at any chemical
potential, the number of quasiparticle excitations equals the number of
bare electrons in the system.  This accounting clearly breaks down in
a real doped Mott system because the number of ways of adding an
electron
is simply $2x$ but the phase space available to add a particle at low energies, namely
$L$, exceeds\cite{harris} $2x$ as shown above.  Hence, the number of low-energy particle addition states per
electron per spin exceeds unity.  As a result in a doped Mott insulator, there are
electronic states at low energy that have no counterpart in the
non-interacting system ($L/n_h>1$) and Fermi liquid theory breaks
down.  

\section{Final Remarks}

We have proposed here that the sign change of the thermopower in the
cuprates is driven by dynamical spectral weight transfer. That our calculated values for the thermopower (see Fig. (\ref{fig2}))
diamonds) agree perfectly with the exact atomic limit for $U=20t$ lends credence
to this interpretation.  The cuprates reside at smaller values of
$U=8t$ where the dynamical mixing with the upper band is significant. Further,  we propose here that close to the top of the dome where particle-hole
symmetry is dynamically generated\cite{hk,jarrell}, a quantum critical
point obtains signalling a transition to weak-coupling physics.  That is, in the strongly overdoped regime, the coupling to the high-energy scale ceases, giving rise to
 $L/n_h=1$, and
traditional descriptions in terms of Landau quasiparticles
apply. Recent soft x-ray Oxygen K-edge experiments\cite{Peets} indicate that
$L/n_h$ does saturate to a doping independent value in the
overdoped regime once the pseudogap terminates as predicted here.
Similar experiments should be performed as a function of temperature
below and above the $T^\ast$ line.  According to the theory we have
recently constructed\cite{ftm0} to explain the onset of non-Fermi
liquid behaviour in a doped Mott insulator, 
$L/n_h$ below $T^\ast$ should increase.  The increase should be given
by $\alpha$, for which quantitative estimates have been obtained recently\cite{shila2}.  The
fundamental problem of superconductivity in the cuprates reduces to
uncloaking precisely how the high-energy scale physics mediated by
dynamical spectral weight transfer creates a phase coherent condensate.

\acknowledgements This work was funded by the NSF DMR-0605769 and DMR-0940992, NSF
PHY05-51164 (KITP) and
Samsung scholarship (S.\ Hong).  We also thank F.
Kr\"uger, T. Stanescu, R. Leigh, and G. Sawatzky for critical remarks.

\end{document}